\documentclass[%
 reprint,
 amsmath,amssymb,
 aps,
superscriptaddress
]{revtex4-2}

\usepackage{graphicx}
\usepackage{dcolumn}
\usepackage{bm}
\usepackage{multirow}
\usepackage{xcolor}
\definecolor{BLUE}{RGB}{0, 0, 255}
\usepackage{adjustbox}

\begin{document}

\title{Ferroelectric Properties of van der Waals Chalcogenides: DFT perspective}

\author{Xue Li}
\affiliation{Department of Physics and Astronomy, University of Manchester, Oxford Road, Manchester, M13 9PL, United Kingdom}
\affiliation{National Graphene Institute, University of Manchester, Booth St.\ E., Manchester, M13 9PL, United Kingdom}
\author{James G. McHugh}
\email{james.mchugh@manchester.ac.uk}
\affiliation{Department of Physics and Astronomy, University of Manchester, Oxford Road, Manchester, M13 9PL, United Kingdom}
\affiliation{National Graphene Institute, University of Manchester, Booth St.\ E., Manchester, M13 9PL, United Kingdom}
\author{Vladimir I. Fal'ko}
\affiliation{Department of Physics and Astronomy, University of Manchester, Oxford Road, Manchester, M13 9PL, United Kingdom}
\affiliation{National Graphene Institute, University of Manchester, Booth St.\ E., Manchester, M13 9PL, United Kingdom}

\begin{abstract}
Layered materials with non-centrosymmetric stacking order are attracting increasing interest due to the presence of ferroelectric polarization, which is dictated by weak interlayer hybridization of atomic orbitals. Here, we use density functional theory modelling to systematically build a library of van der Waals chalcogenides that exhibit substantial ferroelectric polarization. For the most promising materials, we also analyse the pressure dependence of the ferroelectric effect and charge accumulation of photo-induced electrons and holes at surfaces and internal twin boundaries in thin films of such materials.
\end{abstract}

\maketitle
\section*{Introduction}
\begin{figure}[h!]
    \centering    
    \includegraphics[width=0.9\columnwidth]{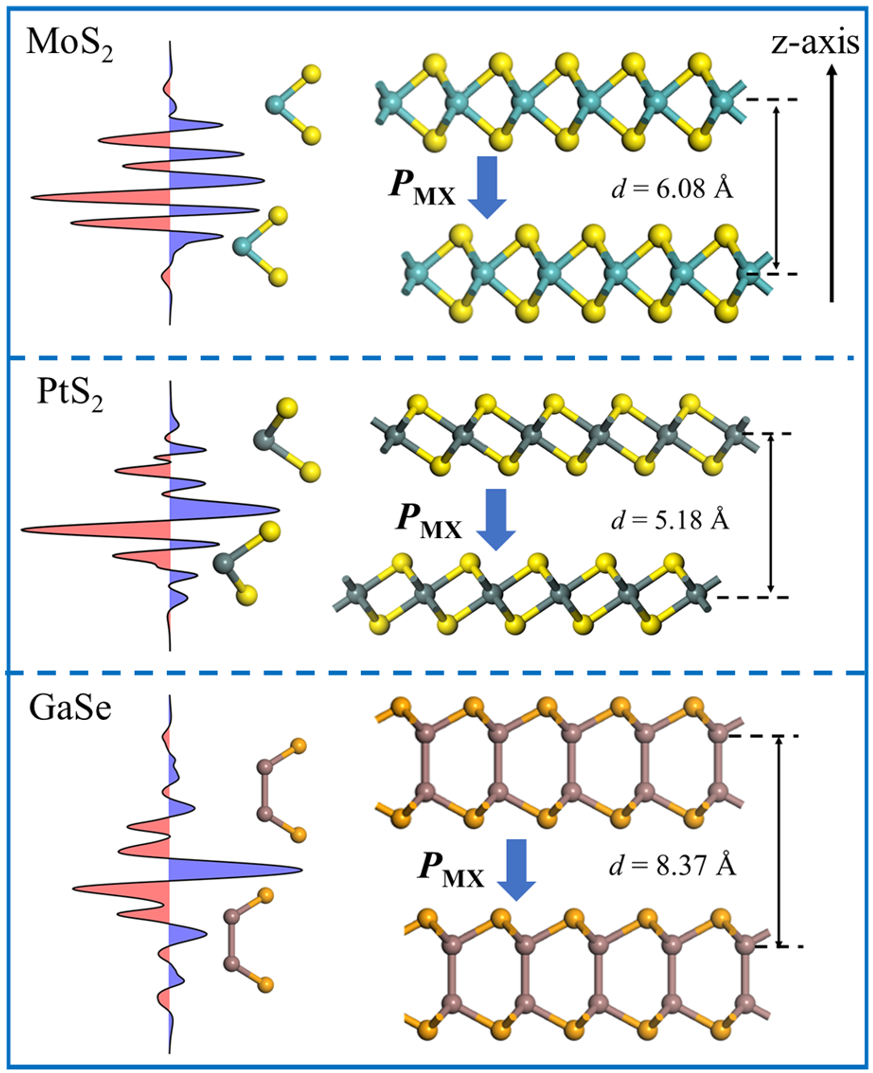}
    \caption{Left: Plane-averaged charge density difference in non-centrosymmetric, MX-stacked bilayers of MoS$_2$, PtS$_2$ and GaSe, demonstrating significant charge redistribution, especially in the interlayer region between the chalcogen atom planes. The red and blue zones represent positive and negative charge density, respectively. Right: Atomic structure of non-centrosymmetric T-MX$_2$, H-MX$_2$ and M$_2$X$_2$ polytypes considered in this work, where blue, grey and brown spheres are metal atoms and $d$ is the interlayer distance. The direction of ferroelectric polarization, $P_{MX}$, arises from interlayer charge transfer.}
    \label{fig:1}
\end{figure}

A remarkable property of certain semiconducting monolayers is their ability to exhibit out-of-plane polarization when stacked into bilayers that lack an inversion center. In general, bulk samples of such materials tend to adopt low-energy stacking configurations that maintain an inversion centre between the layers. In contrast, stacking polymorphs, which lack inversion symmetry, enable asymmetric hybridization between valence and conduction bands of adjacent layers \cite{ferreira2021weak}, leading to interlayer charge transfer $\delta \rho(z)$ and an associated intrinsic electronic polarization, $P = \int z \delta \rho (z)dz$, as shown in Fig. \ref{fig:1}. Since this charge transfer arises from transverse shifting of adjacent layers, rather than a traditional displacive mechanism, these unconventional, layered ferroelectrics have a number of applications, such as transistors \cite{yang2024ferroelectric}, non-volatile memories \cite{gao2024tunnel, martin2016thin}, and novel optoelectronic components \cite{lopez2013ultrasensitive, butler2013progress}. They also allow for ultrafast, fatigue-resistant sliding ferroelectricity \cite{ViznerStern2021,yasuda2024ultrafast}, and moir\'e ferroelectrics \cite{weston2022interfacial,wang2022interfacial}, which can host polar domains with non-trivial topology \cite{bennett2023polar}.

Despite this growing interest, few systematic studies, aimed at identifying improved candidates for the experimental realization of stacking-engineered ferroelectricity, have been conducted, leaving many prospective constituent monolayers unexplored. Here, we analyse the strength of interfacial ferroelectricity in layered van der Waals compounds of metals and chalcogens \cite{Munkhbat2020, Kim2022, zhang2018introduction}, focusing on homobilayers, to facilitate the identification of fundamental physical trends. We consider three classes of these materials, delineated according to lattice structure of constituent monolayers (see Fig. \ref{fig:1}). Transition metal dichalcogenides (TMDs, with chemical unit formula MX$_2$) are the most widely studied and exhibit significant polytypism. The most commonly studied TMDs have a ground state structure with trigonal prismatic coordination of metal atoms in the central sublattice (which we will refer to as H-MX$_2$) and are typically semiconductors. Another common TMD polytype, T-MX$_2$, features octahedral coordination of metal atoms and is related to the H-MX$_2$ polytype by a collective glide of atoms within a chalcogen sublayer. We also consider simple metal chalcogenides (MCs), with stoichiometric formula M$_2$X$_2$.

\begin{figure*}
    \centering    
    \includegraphics[width=2\columnwidth]{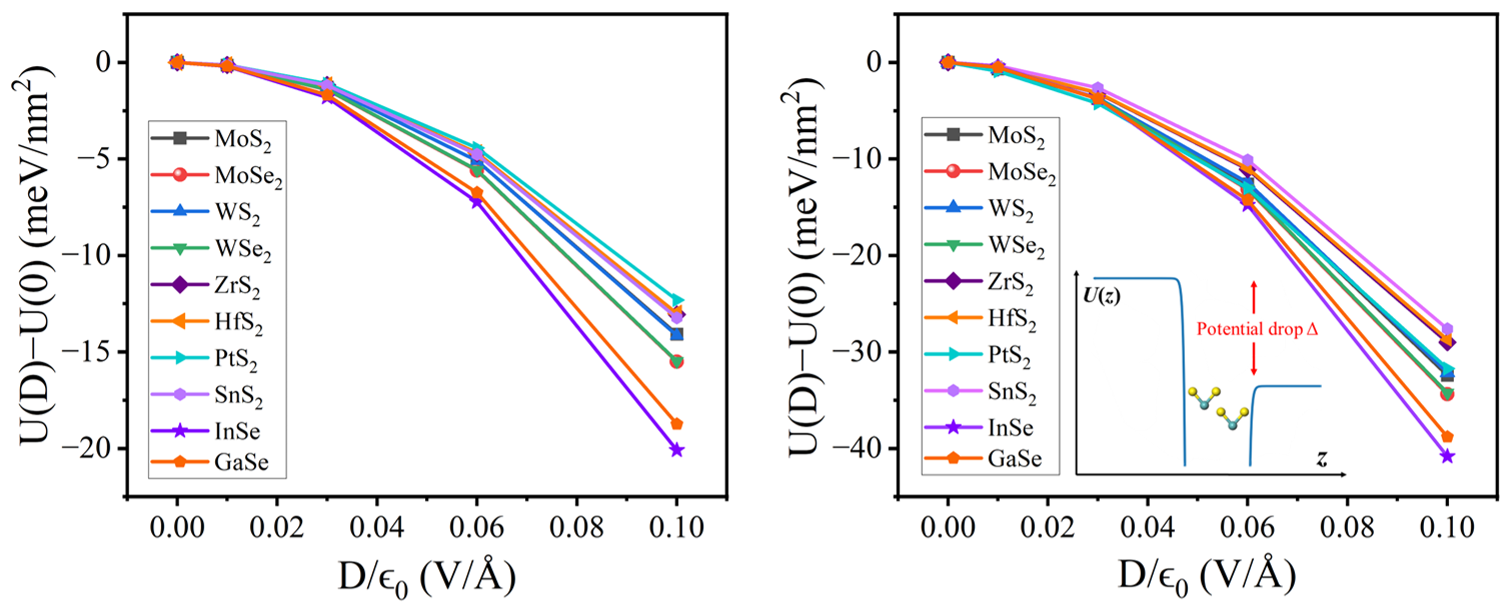}
    \caption{Energy density of monolayers (left: $U(D)-U(0)=-\frac{1}{2}\frac{D^2\alpha^{1L}_{\rm zz}}{\varepsilon_{0} A}$) and MX-stacked bilayers (right: $U(D)-U(0)=-\frac{PD}{\varepsilon_{0}} -\frac{1}{2} \frac{D^2 \alpha^{2L}_{\rm zz}}{\varepsilon_{0} A}$) of H-MX$_2$, T-MX$_2$ and M$_2$X$_2$ as a function of an out-of-plane displacement field ($D$). For bilayers, a linear term is introduced in the fitting equation to account for the contribution of the spontaneous interfacial polarization ($P$). $U(0)$ is the energy density for $D=0$. The inset figure shows the potential drop across the MX bilayer.}
    \label{fig:U}
\end{figure*}

In particular, we examine the out-of-plane polarization ($P$) and the polarizability of monolayers and bilayers ($\alpha^{1L}_{\rm zz}$ and $\alpha^{2L}_{\rm zz}$, which determine the out-of-plane dielectric constant $\varepsilon_{zz}$) of several compounds listed in Table \ref{tab:properties}. These compounds were selected according to the criterion that they are all gapped semiconductors in their bulk form. Parameters are determined from the variation of energy density, computed using density functional theory (DFT) for monolayers (1L) and bilayers (2L) as a function of interlayer distance ($d$) and displacement (electric) field ($D$), analysed using the following approach \cite{PhysRevB.106.125408}, as: 
\begin{equation}
    U(D, \delta) - U(0, 0) = 
    \begin{cases}
    - \frac{1}{2} \frac{D^2 \alpha^{1L}_{\rm zz}}{\varepsilon_{0} A} & \text{1L} \\
        -\frac{PD}{\varepsilon_{0}} -\frac{1}{2} \frac{D^2 \alpha^{2L}_{\rm zz}}{\varepsilon_{0} A} + C \delta^2 & \text{2L}
    \end{cases}
\label{eq1}
\end{equation}
Here, $C$ is the out-of-plane elastic stiffness of van der Waals bilayers, $\delta$ is variation of the bilayer interlayer distance from its equilibrium value, $U(0, 0)$ is the energy per area of a monolayer/bilayer for $D, \delta = 0$, $A$ is the area of unit cell of the 2D crystal and $\varepsilon_{0}$ is the permittivity of vacuum. 

In addition to the above approach, we use a second method to calculate the value of interfacial polarization, $P$, in bilayers \cite{enaldiev2022scalable, meyer2001ab}: we find it from the double layer potential drop, $\Delta$, as $P=\varepsilon_{0} \Delta$. As we find that both methods produce consistent results, we employ the second method to study the dependence of interlayer polarization on pressure applied to bilayers and 3R crystals. For this, we compute the dependence of the potential drop on interlayer spacing, $\Delta (\delta)$, and use elastic stiffness, $C$, to convert the variation of interlayer distance into pressure. 

This report is structured as following. In Section I, we provide details of the performed DFT calculations. In Section II, we use Eq. \ref{eq1} to compute polarizability ($\alpha_{zz}$), polarization ($P$) and stiffness ($C$) from DFT data, analyse the dielectric susceptibility ($\varepsilon_{zz}$) and pressure dependence of the ferroelectric (FE) polarization. In Section III, we estimate the electron density that can be accumulated at the planes of intrinsic twin boundaries inside bulk ferroelectric materials from the list in Table \ref{tab:properties}. 

\begin{figure*}
    \centering  \includegraphics[width=2\columnwidth]{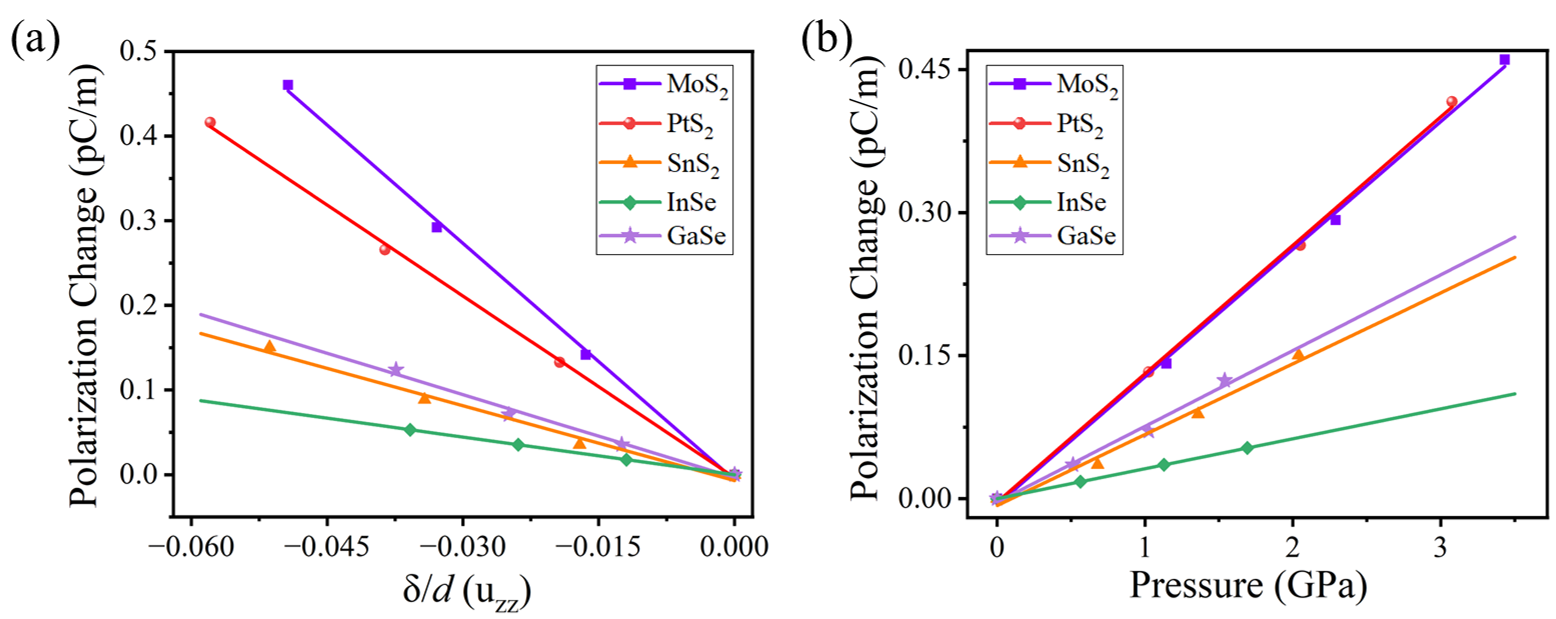}
    \caption{{The pressure effect in MX-bilayers under out-of-plane (a) strain and (b) pressure, which are quantified by the linear slope of the strain, $e_{zzz} = -\frac{\partial P_z}{\partial u_{zz}}$, 
    and pressure dependences, $d_{zzz} = \frac{\partial P_z}{\partial \sigma_{zz}}$.}}
    
    \label{fig:3pressure}
\end{figure*}

\section{DFT calculations}
DFT calculations were performed using Quantum ESPRESSO (QE) \cite{giannozzi2009quantum, giannozzi2017advanced, giannozzi2020quantum}. Plane-wave kinetic energy cutoffs of 80 Ry for wavefunctions and 800 Ry for the charge density, as well as a $22 \times 22 \times 1$ uniform Monkhorst-Pack {\it k}-point sampling of the Brillouin zone, were found to give sufficiently converged results. The thickness of the vacuum layer is set as 15 Å to minimize unphysical interaction between periodic interlayers.

The generalized gradient approximation (GGA) of the exchange correlation functional was employed, as parameterised according to Perdew-Burke-Ernzerhof (PBE) \cite{boese2000new, he2014accuracy, fabiano2010generalized}, and van der Waals effects were incorporated by applying the c09 correction (vdW-DF2-c09), which gives particular accuracy in accounting for interlayer dispersion forces \cite{thonhauser2015spin, langreth2009density, berland2015van, langreth2009density, sabatini2012structural}. During structural relaxation, ultrasoft pseudopotentials were employed to approximate the interactions between nuclei and electrons \cite{schwerdtfeger2011pseudopotential, kresse1999ultrasoft, milman2000electronic}. Structural relaxation of atomic positions and in-plane lattice parameters (while the out-of-plane lattice parameter was fixed at a constant vacuum separation), was performed using the Broyden–Fletcher–Goldfarb–Shanno (BFGS) quasi-Newton algorithm \cite{nawi2006improved, hu2006bfgs}. Non-collinear spin–orbit coupling was incorporated into all self-consistent field (SCF) and band structure calculations using ultra-soft, fully relativistic pseudopotentials \cite{hamann2013optimized}. The interlayer potential drop, $\Delta$, was computed from the plane-averaged local electrostatic potential (including ionic and Hartree contributions) following an initial SCF calculation with an applied dipole correction.

Atomic positions and lattice constants for H-MX$_2$ (MoS$_2$, MoSe$_2$, WS$_2$, WSe$_2$), T-MX$_2$ (ZrS$_2$, ZrSe$_2$, HfS$_2$, HfSe$_2$, PdS$_2$, PdSe$_2$, PtS$_2$, PtSe$_2$, PtTe$_2$, SnS$_2$) and M$_2$X$_2$ (InSe, GaSe) were first sourced from the Computational 2D Materials Database (C2DB) \cite{haastrup2018computational, gjerding2021recent}. These initial configurations were structurally relaxed, giving lattice constants which were in good agreement with C2DB values, differing by no more than 0.05 \text{\AA}. The computed monolayer band structures are also compared with the corresponding data from the C2DB, and we again find consistent values, with a maximum difference between monolayer band gaps of 0.1 eV.

From these monolayer structures, bilayers were constructed in two different orientations, which we label as parallel (P) and anti-parallel (AP). For P-stacked bilayers, both monolayers have a relative orientation of $0^{\circ}$, whereas AP-stacked bilayers have an angle of $180^{\circ}$ between the adjacent monolayers. In each orientation, different stackings are then created by in-plane translations, and are labelled according to the relative position of the interlayer atoms (i.e. those atoms which are closest to the adjacent layer). For H-MX$_2$ and M$_2$X$_2$, only P-bilayers exhibit ferroelectric polarization, whereas for T-MX$_2$ (where individual monolayers have inversion symmetry), ferroelectricity is restricted to AP-bilayers. For each of those two realisations, the possible high-symmetry stackings are labeled as MX (metal-over-chalcogen), XM (chalcogen-over-metal), XX (chalcogen-over-chalcogen), 2H (metal-over-chalcogen on both sublattices) and MM (metal-over-metal), as illustrated in SI Fig. \ref{fig:S1}.  

The binding, quantified as interlayer cohesive energy, 
\begin{equation}
    U_{\rm c}=U_{\rm 2L}-2U_{\rm 1L},
\label{eq2}
\end{equation} 
is then analysed for each stacking to determine its structural stability, where U$_{\rm 2L}$ and U$_{\rm 1L}$ are the energy per cell of bilayer and monolayer, respectively. The calculated values of binding energy, in-plane lattice parameter ($a$) and out-of-plane interlayer spacing ($d$) for the lowest-energy AP and P stacking for each material are given in SI Table \ref{tab:structure}. The obtained value of binding energy in all stacking configurations is negative, indicating that the combination of two monolayers is energetically favourable. For every material, we find that centrosymmetric bilayers, with 2H (H-MX$_2$, M$_2$X$_2$) or MM (T-MX$_2$) stacking, have the strongest cohesive energy. Non-centrosymmetric stacking is systematically, slightly less favourable, with two lowest-energy MX/XM-stacking configurations which are out-of-plane inversion partners. This is consistent with the general tendency of bulk TMDs to crystallize in centrosymmetric structures, while non-centrosymmetric bulk can generally be grown depending on thermodynamic conditions. Before further analysis, we then calculated the bulk band structure of the lowest-energy centrosymmetric stacking. For a number of materials, it was found that interlayer hybridization closed the gap, producing metallic or semi-metallic behaviour; in the following we disregard such materials and focus on those which maintain a gap either in the bulk or at least in their bilayer form.

\begin{table*}[ht]
\centering
\setlength{\tabcolsep}{14pt}
\begin{tabular}{|c|c|c|c|c||c|c||c|c|}
\hline
 & {\bf $P$}/{\bf $P_{\Delta}$} & {\bf $\Delta$} & {\bf $\alpha^{1L}_{zz}$}  & {$\alpha^{2L}_{zz}$} &  {\bf$\varepsilon_{zz}$} & {\bf $E_{FE}$}  & {\bf $C$} & {\bf $d_{zzz}$}   \\
& [$pC/m$] & [$mV$] &  [$\text{\AA}^{3}$] & [$\text{\AA}^{3}$] &   &  [mV/$\rm{\AA}$] &  [N/m] & [pm/V]  \\
\hline
MoS$_2$ & 0.620/0.622 & 70 & 43.50 & 88.89 & 5.96 & 11.51 & 3.52 & 2.49  \\
WS$_2$ & 0.593/0.593 & 67 & 44.16 &  88.88 & 6.05 & 10.93 & 3.45 & 2.59  \\
MoSe$_2$ & 0.522/0.523 & 59 & 52.12 & 105.141  & 7.57 & 9.16 & 3.43 &  2.21 \\
WSe$_2$ & 0.505/0.510 & 57 & 52.23 & 104.96 & 7.24 & 8.78 & 3.51 & 2.26 \\
\hline
GaSe & 0.203/0.201 & 23 & 82.85 & 166.38 & 6.47 & 2.87 & 2.30 & 1.36 \\
InSe & 0.053/0.056 & 6 & 101.16 & 203.88 & 7.63 & 0.72 & 2.91 & 0.06 \\
\hline
PtS$_2$ & 1.062/1.059 & 120 & 48.17 & 98.81 & 7.18 & 23.17 & 4.00 & 1.79  \\
ZrSe$_2$ & 0.803/0.797 & 90 & 64.03 & 129.25 & 6.32 & 14.42 & 4.83 & 0.84 \\
HfSe$_2$ & 0.722/0.726 & 82 & 62.43 & 125.32 & 6.10 & 13.14 & 4.25 & 0.86 \\
ZrS$_2$ & 0.443/0.428 & 50 & 54.29 & 109.38 & 5.16 & 8.53 & 3.60 &  0  \\
HfS$_2$ & 0.416/0.419 & 47 & 52.81 & 105.90 & 5.07 & 8.02 & 3.83 & 0.37 \\
SnS$_2$ & 0.115/0.114 & 13 & 55.45 & 112.98 & 5.55 & 2.23 & 2.94 &  0.92 \\
\hline
\hline
PtSe$_2$ & -/0.859 & 97 & - & - & -& - & -  & - \\
PdS$_2$ & -/0.708 & 80 & - & - & -& - & -  & - \\
PdSe$_2$ & -/0.398 & 45 & - & - & - & - & -  & - \\
PtTe$_2$ & -/0.319 & 36 & - & - & - & - & -  & - \\
\hline
\end{tabular}
\caption{Spontaneous ferroelectric polarization, $P$ (fitted from Eq. \ref{eq1}) \textit{vs} $P_\Delta =\varepsilon_{0} \Delta$; interlayer potential drop, $\Delta$; out-of-plane polarizability ($\alpha^{1L}_{\rm zz}$ for monolayer and $\alpha^{2L}_{\rm zz}$ for bilayer); dielectric constant, $\varepsilon_{zz}$ estimated from polarizability using Eq. \ref{eq3}; intrinsic ferroelectric field, $E_{FE}$; elastic stiffness, $C$; and out-of-plane pressure response ($d_{zzz}$) in MX stacking configurations. Upper section of this table contains semiconductors, sorted by polytype (H-MX$_2$, M$_2$X$_2$ and T-MX$_2$); bottom lines are for metallic T-MX$_2$, so that the results applicable only to bilayers.}
\label{tab:properties}
\end{table*}

\section { Dielectric response and interlayer ferroelectricity}

Using the DFT-computed energies, as displayed in Fig. \ref{fig:U}, we use Eq. \ref{eq1} to determine out-of-plane polarizability and polarization of monolayers and bilayers, by fitting to the obtained values. We note, previous work has shown that lattice relaxation only makes a negligible contribution towards the total polarizability, $\alpha_{zz}$, in 2H-TMDs \cite{PhysRevB.106.125408}. We therefore employ a clamped-ion approach in evaluating all dielectric properties in our calculations. Our results are collected in Table \ref{tab:properties}, and are in agreement with previous works for MoS$_2$, WS$_2$, WSe$_2$ and MoSe$_2$ \cite{ferreira2022scaleability}. Obtained values of bilayer polarizability $\alpha^{2L}_{\rm zz}$ are roughly twice those of the monolayer. We observe that the polarization ($P$) values derived from the two approaches described in the Introduction agree closely, indicating that either method reliably calculates the polarization in the bilayer system. Among all materials listed in Table \ref{tab:properties}, PtS$_2$ clearly stands out, demonstrating a large value of the polarization, with an interlayer potential drop $\Delta \approx 120$ meV, which offers new opportunity for ferroelectric twistronic applications, both in bilayers and thicker films. With reference to the data shown in Table \ref{tab:properties}, we also note a relatively large polarization in bilayers of PtSe$_2$, corresponding to $\Delta \approx 96$ meV, however, in the bulk form this material was reported as semimetallic \cite{villaos2019thickness}, so that its ferroelectric properties would be noticeable only in thin films. 

We also quantify the out-of-plane dielectric susceptibility ($\varepsilon_{zz}$) of bulk TMD and MC crystals, which is largely dictated by the electronic polarizability of the constituent layers, and may be evaluated from monolayer and bilayer calculations due to linear scalability versus the number of layers \cite{ferreira2022scaleability}. The electric field within a bulk crystal consists of two components: external electric field ($D/\varepsilon_{0}$) and the field induced by the polarized charges ($\alpha^{1L}_{\rm zz}$) at the surfaces \cite{ferreira2022scaleability}. Consequently, the following expression can be employed to estimate the $\varepsilon_{zz}$ values
\begin{equation}
    \varepsilon_{zz} = ({1- \frac{\alpha^{1L}_{\rm zz}}{Ad}})^{-1},
\label{eq3}
\end{equation}
where $d$ is the interlayer spacing. All the resulting values estimated in this work are summarised in Table \ref{tab:properties}.

\begin{figure*}[ht]
    \centering    
    \includegraphics[width=2\columnwidth]{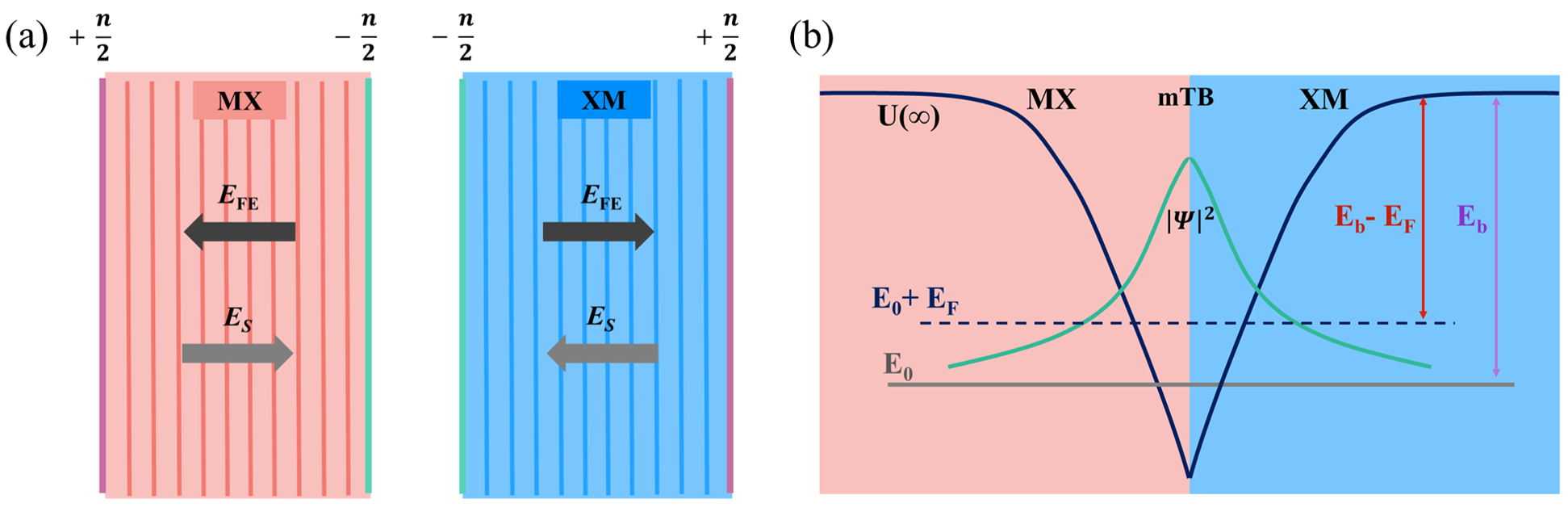}
    \caption{{ (a) An accumulation of opposite charges on the surfaces of MX (red line) and XM (blue line) stackings generates a screening field $E_S$ that cancels the built-in ferroelectric field $E_{FE}$. The green and purple lines are n-type and p-type mTBs, respectively. (b) Quantum well depth $U(\infty)$, ground state energy $E_{0}$,  binding energy $E_{b}$ and Fermi energy $E_{\rm F}$ for electrons and holes at the mTBs and surfaces, and the green curve displays the probability density ($|\Psi(z)|^2$}) of accumulated carriers. The blue curve is the potential profiles in triangular well. }
    \label{fig:density}
\end{figure*}

Finally, we examine how much pressure and strain could change the polarization of these materials. For this, we apply an out-of-plane compression between the consecutive layers and calculate the resulting change in energy per area and potential drop, $\Delta$, to evaluate the pressure response, following the scheme described in Section I. The interlayer distance between the constituent layers of an MX bilayer are reduced in the range of $\delta = 0-0.3$ $\text{\AA}$ in increments of 0.1 $\text{\AA}$. The change in energy per area of the bilayer is calculated for each value and used to extract the 2D out-of-plane elastic constant through the application of Hooke's law. The resulting linear pressure effect can be understood as a first-order coupling between surface polarization ($P$) and the strain ($u_{zz} = \delta/d$) or pressure ($\sigma_{zz}$). It can be characterized by the $e_{zzz}$ and $d_{zzz}$ components of the pressure response tensors, which are also related through the parameter $C$. We then calculate $e_{zzz}$ directly from the linear fitting of polarization change \textit{vs} out-of-plane strain (see Fig. \ref{fig:3pressure}):
\begin{gather}
\notag    U(\delta) = C \delta^2, \quad
       \sigma_{zz}=- \frac{\partial U(\delta)}{\partial \delta}=-2C\delta, \\
    e_{zzz} = -\frac{\partial P_z}{\partial u_{zz}}, \quad d_{zzz} \equiv  \frac{\partial P_z}{\partial \sigma_{zz}} = \frac{e_{zzz}}{C}.
\label{eq4}
\end{gather}

Calculated values of $e_{zzz}$ are shown in Table \ref{tab:properties}, which we note are in strong agreement with experimental measurements of the strain-response in MoS$_2$/WS$_2$ 
\cite{rogee2022ferroelectricity} and InSe \cite{sui2024atomic}. All of the studied materials show a relatively weak response of polarization to pressure, in particular, we estimate that the amount of pressure subjected due to capacitive forces in gated structures would result in less than 1\% variation in the value of polarization, for all ferroelectric van der Waals chalcogenides studied in this work.

\section{Charge accumulation layers at twin boundaries in bulk van der Waals ferroelectrics}

Due to the built-in ferroelectric fields, thin films of the studied materials have the potential to separate photoexcited electrons and holes to opposite surfaces \cite{mchugh2024two}. For films composed of several domains (multilayer twins) of different FE polarization, mirror twin boundary (mTB) interfaces between these domains, also may act as sinks for charge carriers. For a material characterized by ferroelectric field $E_{FE} = \Delta/d$, the maximum charge accumulated at single twin boundary is \cite{mchugh2024two}, 
\begin{equation}
    n = 2E_{\rm FE}\varepsilon_{zz}\varepsilon_{0}/e,
\label{eq5}
\end{equation}
whereas external surfaces would hold charge accumulation of $n/2$, see Fig. \ref{fig:density}. The corresponding values of electron/hole density are collected in Table \ref{tab:binding}, cataloguing  all FE semiconductors we analyzed in this study. 

For each of these expected accumulation layers, we investigated the stability of charge carriers trapped by the corresponding quantum well. That is, we self consistently computed the density profile, and associated quantum well potentials, accumulated at these interfaces/surfaces, within the Thomas-Fermi approach, and, then, compared the obtained binding energies to Fermi energies of electrons and holes with density described by Eq. \ref{eq5}.

\begin{table*}[]
\centering
\setlength{\tabcolsep}{12pt}
\begin{tabular} {|c|c|c|c|c|c|c|}
\hline
 &   & $\ell$ & $g$ & $n$ & mTB $E_{b}$-$E_{F}$ &  surface $E_{b}$-$E_{F}$  \\
&   & [nm]  & & [$\times10^{13}$ ${\rm cm}^{-2}$] & [meV] &  [meV] \\
\hline
\multirow{2}{*}{MoS$_2$\cite{mchugh2024two}} & e & 1.09 & 6 &\multirow{2}{*} {0.76} & 24.0 & 24.0 \\
& h & 0.91 & 2 & & 0.70  & 11.5 \\
\hline
\multirow{2}{*}{MoSe$_2$\cite{mchugh2024two}} & e & 1.18 & 6 & \multirow{2}{*} {0.80} & 19.3 &  20.1 \\
& h & 0.86 & 2 & & 0  & 7.5 \\
\hline
\multirow{2}{*}{WS$_2$\cite{mchugh2024two}} & e & 1.13 & 6 & \multirow{2}{*} {0.73} & 22.6 &  23.3 \\
& h & 0.91 & 2 & & 1.2  & 9.1 \\
\hline
\multirow{2}{*}{WSe$_2$\cite{mchugh2024two}} & e & 1.25 & 6 & \multirow{2}{*} {0.70} & 18.9 &  20.3 \\
& h & 0.91 & 2 & & -0.9  & 7.5 \\
\hline\hline
\multirow{2}{*}{ZrS$_2$} & e & 0.92 & 6 & \multirow{2}{*} {0.49} & 14.7 &  14.88 \\
& h & 1.57 & 2 & & 35.02  & 30.41 \\
\hline
\multirow{2}{*}{ZrSe$_2$} & e & 0.90 & 6 & \multirow{2}{*} {1.01} & 16.40 & 20.96 \\
& h & 1.46 & 2 & & 19.83 & 30.22 \\
\hline
\multirow{2}{*}{HfS$_2$} & e & 0.88 & 6 & \multirow{2}{*} {0.45} & 13.17 &  13.37 \\
& h & 1.56 & 2 & & 32.49  & 28.30 \\
\hline
\multirow{2}{*}{HfSe$_2$} & e & 0.87 & 6 & \multirow{2}{*} {0.89} & 15.92 &  18.93 \\
& h & 0.67 & 2 & & -82.65  & -32.81 \\
\hline
\multirow{2}{*}{PtS$_2$} & e & 0.98 & 6 & \multirow{2}{*} {1.84} & -23.31 &  10.27 \\
& h & 0.99 & 2 & & -29.62  & 7.33 \\
\hline
\multirow{2}{*}{SnS$_2$} & e & 1.49 & 6 & \multirow{2}{*} {0.14} & 6.37 &  6.35 \\
& h & 2.40 & 6 & & 13.66  & 11.93 \\
\hline\hline
\multirow{2}{*}{InSe} & e & 4.95 & 2 & \multirow{2}{*} {0.06} & 0.81 & 3.90  \\
& h & 5.11 & 2 & &  9.85 & 8.53 \\
\hline
\multirow{2}{*}{GaSe} & e & 1.17 & 6 & \multirow{2}{*} {0.21} & 6.13 & 6.44 \\
& h & 2.33 & 2 & &  12.69 & 13.1 \\
\hline
\end{tabular}
\caption{{The difference of binding energies and Fermi energy ($E_b - E_F$), characteristic length of confined states ($\ell$), maximum 2D carrier density ($n$), spin degeneracy factor ($g$) for electron (e) and hole (h) at mTBs and surfaces.}}
\label{tab:binding}
\end{table*}

For all materials with different parameters (band masses, $E_{\rm FE}$, dielectric susceptibilities) the results are drawn from the following set of equations and boundary condition, $\partial_{\xi}f(0)=0$, combining dimensionless wavefunctions, $f(\xi)=\sqrt{\ell}\psi(\ell\xi)$, and  potential, $U(\xi)$, dependent on a dimensionless out-of-plane coordinate, $\xi=\frac{z}{\ell}$, 
\begin{gather}\label{EqDimenless}
    \bigg[ -\frac{1}{2}\partial_{\xi}^{2} + U(\xi) \bigg] f(\xi) = u f(\xi), \quad \xi\geq 0,\\
    \notag U(\xi) = \xi - 2 \int_{0}^{\xi}d\xi' \, (\xi-\xi') |f(\xi')|^{2},\\
    \notag f(\infty) = 0, \quad \int_{0}^{\infty}d\xi\, |f(\xi)|^{2} = \frac{1}{2}.
\label{eq6}
\end{gather}
Numerical solution of this set of equations determines the binding energy of charge carriers, $E_b=\frac{[U(\infty)-u] \hbar^{2}}{m_{zz}\ell^2}$, where $\ell=\bigg(\frac{\hbar^{2}}{em_{z} E_{\rm FE}}\bigg)^{\frac{1}{3}}$ and $U(\infty)-u = 0.271$. Then, information about the accumulation layer stability is given by the difference, $E_{b}-E_{F}$, between the binding and Fermi energies of the 2D electron gas, $E_F = \frac{2n \pi \hbar^{2}}{\sqrt{m_{x}m_{y}}g}$, where $g$ is the degeneracy factor determined by the conduction/valence band edge properties of each material, and $m_{x, y}$ are the corresponding in-plane effective masses.

All the resulting material-specific parameters for the accumulation layers are displayed in Table \ref{tab:binding}. The upper part of this table reproduces the earlier-established \cite{mchugh2024two} stability criteria for H-MX$_2$. Among the newly studied T-MX$_2$, PtS$_2$ offers no stability for either electrons or holes, mostly, due to their light in-plane masses, whereas all other semiconducting FE materials would have stable accumulation layers at least, at low temperatures. In addition, Table \ref{tab:binding} contains information about stability of surface accumulation layers (Fig. \ref{fig:density}a). The latter were found by solving \cite{mchugh2024two} the set of Eqs. 6 with a boundary condition, $f(0)=0$, instead of $\partial_{\xi}f(0)=0$, and the results indicate that surface accumulation layers are capable of holding carrier density $n/2$ for all studied FE materials.

\section*{Conclusions}
Overall, we overview the broad range of van der Waals materials that feature interfacial ferroelectricity. In addition to the earlier-established group of four semiconducting transition metal dichalcogenides with honeycomb layer structure, here, we identify several FE candidates among tetrahedral transition metal dichalcogenides. In fact, the strongest ferroelectric polarization appears to be in PtS$_2$, with built-in ferroelectric field about twice larger than in other studied materials, and an associated potential drop comparable to P-stacked hBN bilayers ($\Delta = 100-110$ mV) \cite{ViznerStern2021, yasuda2021stacking}. This offers new materials prospects for building ferroelectric devices which have reversible out-of-plane polarisation, switchable by interlayer sliding. Similarly to bilayers of hBN and 2H-TMDs, this will also allow the creation of superlattices of triangular stacking domains with alternating values of $P$, through the application of an interlayer twist \cite{weston2022interfacial}. In addition, the functional properties of such bilayers can be further enriched, for example through the incorporation of dopant atomic species, which can enhance intrinsic ferroelectricity \cite{cardenas2023room} or induce multiferroicity \cite{sui2024atomic}.

We further demonstrate that ferroelectric parameters of these materials are almost insensitive to pressure, at least for the pressures induced by capacitive effects in gated structures across the voltage providing displacement fields needed for FE polarization reversal. Finally, we also remark that we find stable charge accumulation at mTBs and external surfaces for all materials studied, apart from mTBs in PtS$_2$, suggesting the possibility of these interfaces to realise distinct correlated states when compared to H-MX$_2$ TMDs \cite{mchugh2024two}.

\bibliography{references}

\begin{widetext}
\newpage
\section{Supplementary Information}
\setcounter{equation}{0}
\renewcommand{\theequation}{E\arabic{equation}}

\setcounter{figure}{0}
\renewcommand{\thefigure}{\arabic{figure}}
\renewcommand{\figurename}{Supplementary Information Fig.}

\setcounter{table}{0}
\renewcommand{\thetable}{\Roman{table}}
\renewcommand{\tablename}{Supplementary Information Table}

\begin{figure}
    \centering    
    \includegraphics[width=0.7\columnwidth]{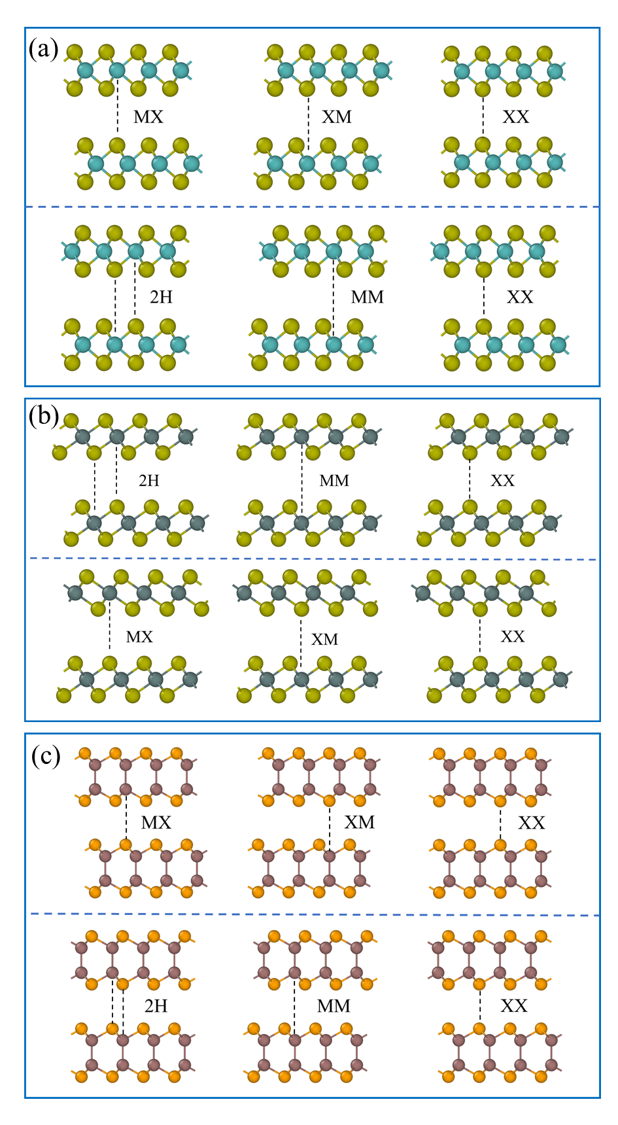}
    \caption{ Atomic structures of non-centrosymmetric (a) H-MoS$_2$, (b) T-SnS$_2$ and (c) MX (top: P bilayers. bottom: AP bilayers), which are the three prototypical classes of monolayers considered in this work. The blue and yellow spheres denote the metal atom chalcogen atom, respectively.}
    \label{fig:S1}
\end{figure}

\begin{table*}[ht]
\centering
\caption{Structural properties and band gaps of all TMD and MCs bulk (in-plane: a, out-of-plane lattice constants: c/2) and out-of-plane interlayer spacing (d) of AP and P stackings, which considered in this work. In all cases, the centrosymmetric stacking of constituent layers is found to give the lowest binding energy, while non-centrosymmetric structures are slightly higher in energy. }
\begin{tabular}{| r|c|c|c|c|c|c|c|c|}
\hline
   &  {\bf a (bulk)} & {\bf c/2 (bulk) } & {\bf d (AP) }  & {\bf d (P) } & {\bf E$_c$ (AP)} & {\bf E$_c$ (P) } & {\bf Band gap (DFT)} & {\bf Band gap (expt)} \\
  &  [$\text{\AA}$] & [$\text{\AA}$] & [$\text{\AA}$] & $\text{\AA}$ &  [eV] & [eV] & [eV] & [eV] \\
\hline
MoS$_2$ & 3.15  & 6.12  & 6.12  & 6.08  & -0.187  & -0.185 & 0.8406 & 1.23\cite{gusakova2017electronic} \\
MoSe$_2$ & 3.28 & 6.47 & 6.45  & 6.44  & -0.200  & -0.196 & 0.8058 & 1.09\cite{gusakova2017electronic} \\
WS$_2$ & 3.16 & 6.15 & 6.15  & 6.13  & -0.184  & -0.181 & 0.9186 & 1.32\cite{gusakova2017electronic} \\
WSe$_2$ & 3.28  & 6.49 & 6.50  & 6.49  & -0.198  & -0.191 & 0.8606 & 1.21\cite{gusakova2017electronic} \\
\hline
ZrS$_2$ & 3.64 & 5.79 & 5.86 & 5.78 & -0.207 & -0.224 & 0.7099 & 1.78\cite{roubi1988resonance}, 1.70\cite{lee1969optical, greenaway1965preparation}  \\
ZrSe$_2$ & 3.75  & 6.11  & 6.24  & 6.10  & -0.205  & -0.230 & -0.0238 & 1.2\cite{lee1969optical}, 1.18\cite{moustafa2009growth}, 1.10\cite{starnberg1996photoemission}  \\
HfS$_2$ & 3.60  & 5.81  & 5.86  & 5.82  & -0.181  & -0.197 & 0.9013 & 1.96\cite{greenaway1965preparation}, 1.87\cite{gaiser2004band}, 1,80\cite{terashima1987indirect} \\
HfSe$_2$ & 3.71  & 6.14  & 6.24  & 6.13  & -0.197  & -0.217 & 0.1445 & 1.15\cite{terashima1987indirect}, 1.13\cite{greenaway1965preparation} \\
PdS$_2$ & 3.54   & 4.64  & 5.05  & 4.64  & -0.242  & -0.293 & 0.0258 & -0.137\cite{feng2020layer} \\
PdSe$_2$ & 3.73  & 4.85  & 5.16  & 4.86  & -0.305  & -0.350 & -0.4191 & 0.4\cite{wilson1969transition} \\
PtS$_2$ & 3.56  & 4.81  & 5.18  & 4.81  & -0.209  & -0.247 & 0.1226 & 1.1\cite{guo1986electronic}, 0.95\cite{tributsch1982photoelectrochemical}, 0.87\cite{parsapour1998spectroscopy}, 0.70\cite{hulliger1965electrical} \\
PtSe$_2$ & 3.73   & 5.05  & 5.38   & 5.05  & -0.243  & -0.273 & 0.2758 & 0\cite{kandemir2018structural, zhao2017high}  \\
PtTe$_2$ & 4.03  & 5.21   & 5.59  & 5.21  & -0.344  & -0.407 & -0.5437 & 0\cite{han2022strong} \\
SnS$_2$ & 3.67  & 5.83  & 5.84  & 5.82 & -0.174  & -0.177 & 1.1674 & 2.28\cite{burton2016electronic} \\
\hline
InSe & 4.03  & 8.23  & 8.31   & 8.37 & -0.198  & -0.194 & 0.209 & 1.30\cite{mudd2016direct}, 1.25\cite{bandurin2017high, wu2019crystal}  \\
GaSe & 3.77  & 7.79  & 7.84 & 8.02 & -0.169  & -0.169 & 0.7256 & 2.12\cite{aulich1969indirect, allakhverdiev2009effective}, 2.01\cite{rahaman2018vibrational}  \\
\hline
\end{tabular}
\label{tab:structure}
\end{table*}


\begin{table*}[]
\centering
\caption{{\textbf{Electron (e) and hole (h) properties at mTBs and surfaces.} Intrinsic FE field $E_{\rm FE}$, maximum 2D carrier density $n$, out-of-plane dielectric constant $\varepsilon_{zz}$, in-plane and out-of-plane effective mass $m_{xy}$ $m_{zz}$, characteristic length of confined states $\ell$. Ground state energy $E_{0}$, quantum well depth $U(\infty)$, binding energy $E_{b}$ and Fermi energy $E_{F}$ for electrons and holes at the mTBs and surfaces.}}
\begin{tabular} {|c|c|c|c|c|c|c|cccc|cccc|}
\hline
 &  & $E_{\rm FE}$ & $n$ & {\bf$\varepsilon_{zz}$} & $m_{xy}$/$m_{zz}$ & $\ell$ &  \multicolumn{4}{c|}{mTB} &  \multicolumn{4}{c|}{surface} \\
 & & & & & & & $E_{0}$ & $E_{b}$ & $U(\infty)$ & $E_{\rm F}$ & $E_{0}$ & $E_{b}$ & $U(\infty)$ & $E_{\rm F}$   \\
 & &  [V/${\rm nm}$]  &  [$\times10^{13}$ ${\rm cm}^{-2}$]  &  & [$m_0$] & [nm] & [meV] & [meV] & [meV] & [meV]  & [meV] & [meV] & [meV]& [meV] \\
\hline
\multirow{2}{*}{MoS$_2$ \cite{mchugh2024two}} & e & \multirow{2}{*}{0.115}  & \multirow{2}{*}{0.8} & \multirow{2}{*}{6.10} & 0.53/0.56/0.51 & 1.09 & 72.4 & 34.0  & 106.4 & 10.0 & 198.8 & 29.0 & 227.8 & 5.0\\
& h &   &   &  & 0.67/0.67/0.88 & 0.91 &  60.4 & 28.4 & 88.7 & 27.7 &  164.8 & 24.1 & 188.9 & 12.6  \\
\hline
\multirow{2}{*}{MoSe$_2$ \cite{mchugh2024two}} & e & \multirow{2}{*}{0.092}  & \multirow{2}{*}{0.7} & \multirow{2}{*}{7.34} & 0.48/0.71/0.50 & 1.18 & 62.8 & 29.5 & 92.3 & 10.2 &  172.4 & 25.2 & 197.6 & 5.1  \\
& h &   &   &  & 0.83/0.83/1.30 & 1.30 & 45.7 & 21.5 & 67.1 & 21.5 &  125.4 & 18.3 & 143.7 & 10.8 \\
\hline
\multirow{2}{*}{WS$_2$ \cite{mchugh2024two}} & e & \multirow{2}{*}{0.110}  & \multirow{2}{*}{0.8} & \multirow{2}{*}{5.84} & 0.53/0.56/0.48 & 1.13 & 71.7 & 33.7 & 105.4 & 11.1 &  196.9 & 28.8 & 225.7 & 5.5  \\
& h &   &   &  & 0.61/0.61/0.75 & 0.91 & 61.8 & 29.1 & 90.9 & 27.9 &  169.7 & 24.8 & 194.5 & 15.7 \\
\hline
\multirow{2}{*}{WSe$_2$ \cite{mchugh2024two}} & e & \multirow{2}{*}{0.092}  & \multirow{2}{*}{0.7} & \multirow{2}{*}{7.20} & 0.45/0.52/0.43 & 1.25 & 65.6 & 30.8 & 96.4 & 11.9 &  180.8 & 26.3 & 206.3 & 6.0  \\
& h &   &   &  & 0.74/0.74/1.10 & 0.91 &  47.9 & 22.5 & 70.5 & 23.4 & 131.6 &  19.2 & 150.9 & 11.7 \\
\hline\hline
\multirow{2}{*}{ZrS$_2$} & e & \multirow{2}{*}{0.085}  & \multirow{2}{*}{0.49} & \multirow{2}{*}{5.16} & 1.23/0.30/1.16 & 0.92 &  45.01 & 21.14 & 66.15 & 6.44 & 123.56 & 18.10 & 141.65 & 3.22  \\
& h &   &   &  & 9.50/9.50/0.23 & 1.57 & 77.18 & 36.25 & 113.44 & 1.23 & 211.89 & 31.03 & 242.92 & 0.62 \\
\hline
\multirow{2}{*}{ZrSe$_2$} & e & \multirow{2}{*}{0.144}  & \multirow{2}{*}{ 1.01 } & \multirow{2}{*}{6.32} & 0.94/0.21/0.73 & 0.90 & 74.71 & 35.09 & 109.79 & 18.69 & 205.09 & 30.04 & 235.13 & 9.08  \\
& h &   &   &  & 0.65/0.65/0.17 & 1.46 & 121.43 & 57.03 & 178.46 & 37.20 & 333.35 & 48.82 & 382.17 & 18.60 \\
\hline
\multirow{2}{*}{HfS$_2$} & e & \multirow{2}{*}{0.080}  & \multirow{2}{*}{0.45} & \multirow{2}{*}{5.07} & 1.45/0.26/1.41 & 0.88 & 40.50 & 19.02 & 59.53 & 5.85 & 111.19 & 16.29 & 127.47 & 2.92  \\
& h &   &   &  & 7.85/7.85/0.25 & 1.56 & 72.10 & 33.86 & 105.96 & 1.37 & 197.92 & 28.99 & 226.91 & 0.69 \\
\hline
\multirow{2}{*}{HfSe$_2$} & e & \multirow{2}{*}{0.131}  & \multirow{2}{*}{ 0.89 } & \multirow{2}{*}{6.10} & 1.20/0.19/0.89 & 0.87 & 65.60 & 30.81 & 96.40 & 14.89 & 180.08 & 26.37 & 206.45 & 7.44  \\
& h &   &   &  & 0.20/0.20/1.91 & 0.67 & 50.85 & 23.88 & 74.74 & 106.53 & 139.61 & 20.45 & 160.05 & 53.26 \\
\hline
\multirow{2}{*}{PtS$_2$} & e & \multirow{2}{*}{0.232}  & \multirow{2}{*}{1.84} & \multirow{2}{*}{7.18} & 0.20/0.15/0.35 & 0.98 & 131.06 & 61.55 & 192.61 & 84.86 & 359.78 & 52.70 & 412.48 & 42.43  \\
& h &   &   &  & 0.48/0.48/0.34 & 0.99 & 132.33 & 62.15 & 194.48 & 91.77 & 363.28 & 53.21 & 416.48 & 45.88 \\
\hline
\multirow{2}{*}{SnS$_2$} & e & \multirow{2}{*}{0.023}  & \multirow{2}{*}{0.14} & \multirow{2}{*}{5.55} & 0.69/0.25/1.04 & 1.49 & 18.96 & 8.90 & 27.86 & 2.53 & 52.04 & 7.62 & 59.66 & 1.27  \\
& h &   &   &  & 1.43/2.80/0.25 & 2.40 & 30.49 & 14.32 & 44.81 & 0.66 &  83.70 & 12.26 & 95.96 & 0.33 \\
\hline\hline
\multirow{2}{*}{InSe} & e & \multirow{2}{*}{0.0072} & \multirow{2}{*}{0.06} & \multirow{2}{*}{7.63} & 0.16\cite{ceferino2020crossover}/0.086\cite{ceferino2020crossover} & 4.95 & 20.85 & 9.79  & 30.65 & 8.98 &  57.25  & 8.39  & 65.64  & 4.49   \\
& h & & & & 5.35\cite{ceferino2020crossover}/0.078\cite{ceferino2020crossover} & 5.11  & 21.54 & 10.12 & 31.66 & 0.27 &  59.15  & 8.66  & 67.80  & 0.13  \\
\hline
\multirow{2}{*}{GaSe} & e & \multirow{2}{*}{0.029}   & \multirow{2}{*}{0.21} & \multirow{2}{*}{6.47} & 0.5\cite{tu2005intraband}/1.6\cite{tu2005intraband} & 1.17  & 20.19 & 9.48  & 29.68 & 3.35 &  55.43  & 8.12  & 63.55  & 1.68 \\
& h & & & & 0.8\cite{tu2005intraband}/0.2\cite{tu2005intraband} & 2.33 & 40.39 & 18.97 & 59.35 & 6.28 & 110.87 & 16.24 & 127.11 & 3.14 \\
\hline
\end{tabular}
\label{tab:details}
\end{table*}

\end{widetext}

\end{document}